\documentclass[letterpaper,reprint, aps]{revtex4-1}
\usepackage[LGR,T1]{fontenc}
\usepackage[latin2]{inputenc}
\usepackage{amsmath}
\usepackage{amssymb}
\usepackage{graphicx}
\usepackage{esint}
\usepackage{subscript}

\makeatletter

\DeclareRobustCommand{\greektext}{%
  \fontencoding{LGR}\selectfont\def\encodingdefault{LGR}}
\DeclareRobustCommand{\textgreek}[1]{\leavevmode{\greektext #1}}
\DeclareFontEncoding{LGR}{}{}
\DeclareTextSymbol{\~}{LGR}{126}

\makeatother

\begin{document}

\title{Complex phase diagram of Ba$_{1-x}$Na$_{x}$Fe$_{2}$As$_{2}$:
a multitude of phases striving for the electronic entropy}

\author{L. Wang}

\email{liran.wang@kit.edu}

\affiliation{Institut f$\ddot{u}$r Festk$\ddot{o}$rperphysik, Karlsruhe Institute
of Technology, 76021 Karlsruhe, Germany}

\author{F. Hardy}

\affiliation{Institut f$\ddot{u}$r Festk$\ddot{o}$rperphysik, Karlsruhe Institute
of Technology, 76021 Karlsruhe, Germany}

\author{A. E. Böhmer}

\email{present address: The Ames Laboratory, U.S. Department of Energy, Iowa State University, Ames, Iowa 50011, USA}

\affiliation{Institut f$\ddot{u}$r Festk$\ddot{o}$rperphysik, Karlsruhe Institute
of Technology, 76021 Karlsruhe, Germany}

\author{T. Wolf}

\affiliation{Institut f$\ddot{u}$r Festk$\ddot{o}$rperphysik, Karlsruhe Institute
of Technology, 76021 Karlsruhe, Germany}

\author{P. Schweiss}

\affiliation{Institut f$\ddot{u}$r Festk$\ddot{o}$rperphysik, Karlsruhe Institute
of Technology, 76021 Karlsruhe, Germany}

\author{C. Meingast}

\email{christoph.meingast@kit.edu}

\affiliation{Institut f$\ddot{u}$r Festk$\ddot{o}$rperphysik, Karlsruhe Institute
of Technology, 76021 Karlsruhe, Germany}

\date{7/12/15}
\begin{abstract}
The low-temperature electronic phase diagram of Ba$_{1-x}$Na$_{x}$Fe$_{2}$As$_{2}$,
obtained using high-resolution thermal-expansion and specific-heat
measurements, is shown to be considerably more complex than previously
reported, containing nine different phases. Besides the magnetic $C_{2}$
and reentrant $C_{4}$ phases, we find evidence for an additional,
presumably magnetic, phase below the usual SDW transition, as well
as a possible incommensurate magnetic phase. All these phases coexist
and compete with superconductivity, which is particularly strongly
suppressed by the $C_{4}$-magnetic phase due to a strong reduction
of the electronic entropy available for pairing in this phase.
\end{abstract}

\keywords{Iron-based superconductor, $C_{4}$-magnetic phase, thermodynamics,
phase diagram}

\maketitle
High-temperature superconductivity in Fe-based systems usually emerges
when a stripe-type antiferromagnetic spin-density-wave (SDW) is suppressed
by either doping or pressure \citep{paglionenature,Johnston:2010aa,Ishida:2009aa}.
The SDW transition is accompanied, or sometimes even slightly preceeded,
by a structural phase transition from a high-temperature tetragonal
($C_{4}$) to a low-temperature orthorhombic ($C_{2}$) state, which
has sparked the lively debate about electronic nematicity and the
respective role of spin and orbital physics in these materials \citep{Kontani2011,Fernandes2012,Fernandes2014,Boehmer:2015aa,Bohmer2015review}.
In the hole-doped compounds, Ba$_{1-x}$Na$_{x}$Fe$_{2}$As$_{2}$,
Ba$_{1-x}$K$_{x}$Fe$_{2}$As$_{2}$, and Sr$_{1-x}$Na$_{x}$Fe$_{2}$As$_{2}$,
recent studies have shown that the $C_{4}$ symmetry is restored in
a small pocket within the magnetic $C_{2}$ phase region \citep{Avci2014,Bohmer:2015ab,Mallett:2015aa,Allred2015}.
Mössbauer studies on Sr$_{0.63}$Na$_{0.37}$Fe$_{2}$As$_{2}$ find
that only half of the Fe sites carry a magnetic moment in this phase
\citep{Allred2015}, which is consistent with the double-Q magnetic
structure predicted within the itinerant spin-nematic scenario \citep{Fernandes2014,Avci2014,Allred2015,Gastiasoro:2015}.
Moreover, neutron studies have shown that the spins flip from in-plane
in the $C_{2}$ phase to out of plane in the $C_{4}$ reentrant phase
\citep{Wasser2015}, indicating that spin-orbit interactions cannot
be neglected. In the Ba$_{1-x}$K$_{x}$Fe$_{2}$As$_{2}$ system,
the reentrant $C_{4}$ phase reverts back to the $C_{2}$ phase near
the onset of superconductivity, due to a stronger competition of the
$C_{4}$ phase with superconductivity \citep{Bohmer:2015ab}. The
presence of this phase in the hole-doped systems presents strong evidence
that the physics of these Fe-based systems can be treated in an itinerant
picture, and recent theoretical studies based upon the spin-nematic
scenario can reproduce phase diagrams very similar to the experimental
ones \citep{Kang2015}, as well as the spin-reorientation in the $C_{4}$
phase if spin-orbit interactions are included \citep{Christensen2015}. 

Here, we reinvestigate in greater detail the low-temperature electronic
phase diagram of Ba$_{1-x}$Na$_{x}$Fe$_{2}$As$_{2}$ using high-resolution
thermal-expansion and specific- heat measurements and show that it
is considerably more complex than previously reported, containing
nine different phases. Besides the usual $C_{2}$ and reentrant $C_{4}$
magnetic phases, we find evidence for an additional, presumably magnetic,
$C_{2}$ phase, in which the orthorhombic distortion is substantially
reduced but still finite. These phases coexist and compete with superconductivity,
which is particularly strongly suppressed by the reentrant $C_{4}$
phase. Further, we provide indications that the SDW transition becomes
incommensurate above x = 0.22, which appears linked to the emergence
of the $C_{4}$ phase at this composition. The surprising occurence
of this multitude of phases near the onset of superconductivity suggests
a highly degenerate free-energy landscape near optimal doping, which
may be related to the occurence of superconductivity in the Fe-based
systems.

\begin{figure*}[tbph]
\centering{}\includegraphics[width=0.7\textwidth]{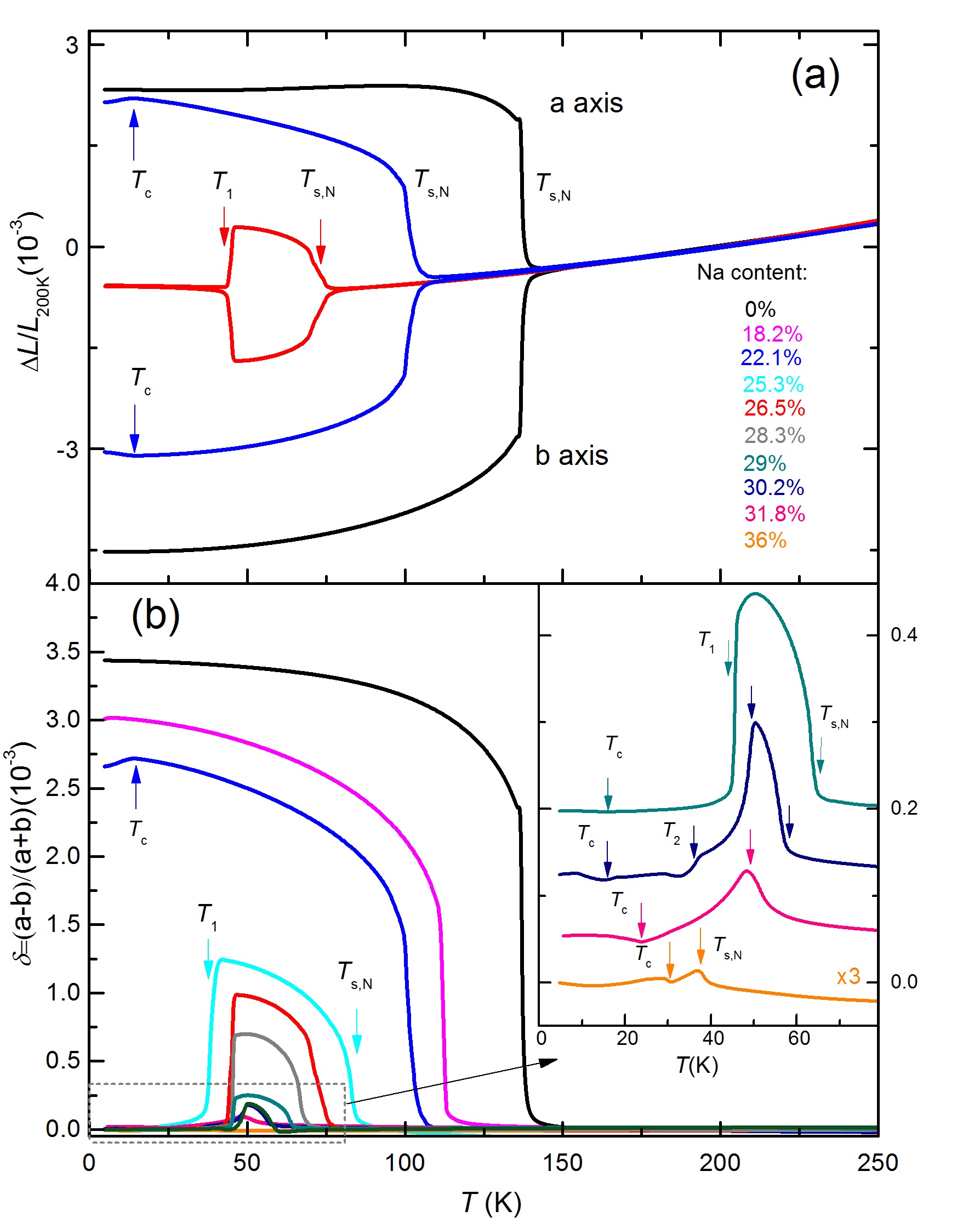} \protect\caption{(a) Relative length change, $\Delta L/L$, versus temperature of the
orthorhombic lattice parameters a and b of Ba$_{1-x}$Na$_{x}$Fe$_{2}$As$_{2}$
for Na doping levels of x = 0, 0.221, 0.265 obtained using high-resolution
capacitance dilatometry (see text for details). (b) Temperature dependence
of the orthorhombic distortion $\delta=(a-b)/(a+b)$ inferred from
the data in (a). The inset presents an expanded view of the data at
higher doping levels. Vertical arrows indicate the location of the
superconducting transition at $T{}_{c}$, the $C_{4}$-reentrant transition
at $T{}_{1}$, and the stripe-type SDW transition at $T{}_{s,N}$.}
\label{Fig1} 
\end{figure*}

\begin{figure*}
\centering{}\includegraphics[scale=0.6]{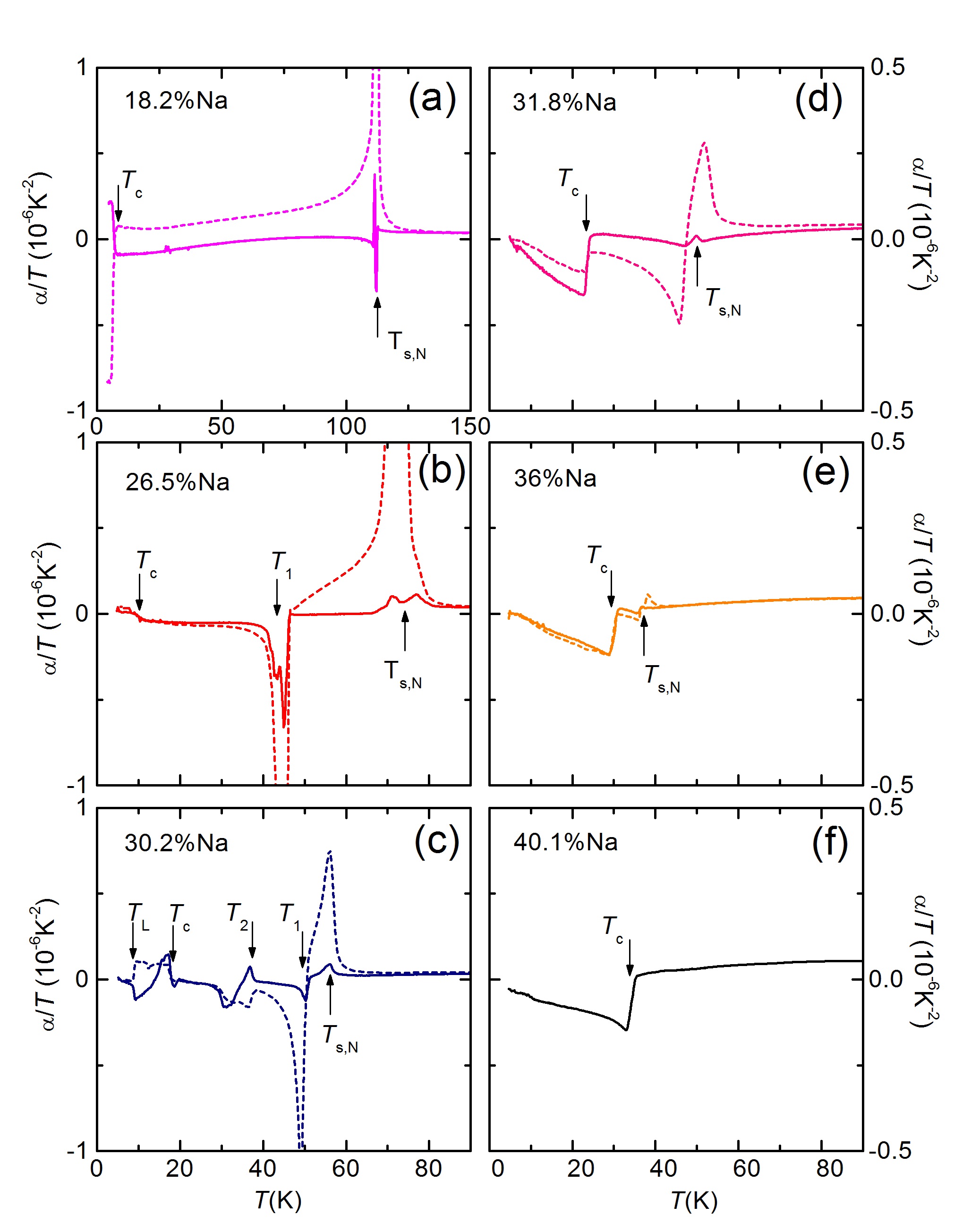}\protect\caption{(a)-(f) In-plane thermal expansion coefficients in 'twinned' (solid
lines) and 'detwinned' (dashed lines) orientations versus T for Na
concentrations of x = 0.182, 0.265, 0.302, 0.318, 0.36, and 0.401.
The location of the various phase transitions is marked by vertical
arrows. The breaking of the C\protect\textsubscript{4} symmetry at
$T_{s,N}$ in (a)-(e) is clearly indicated by the anisotropy of the
'twinned' and detwinned' expansion coefficients below $T_{s,N}$.
On the other hand, the reentrant C\protect\textsubscript{4} phase
is characterized by equivalent expansion coefficients below $T_{1}$
in (b) and between $T_{2}$ and $T_{c}$ in (c). The near optimally
doped sample in f) exhibits only a well-defined jump at $T_{c}$.
\label{Fig2}}
\end{figure*}

\begin{figure*}
\begin{centering}
\includegraphics[scale=0.4]{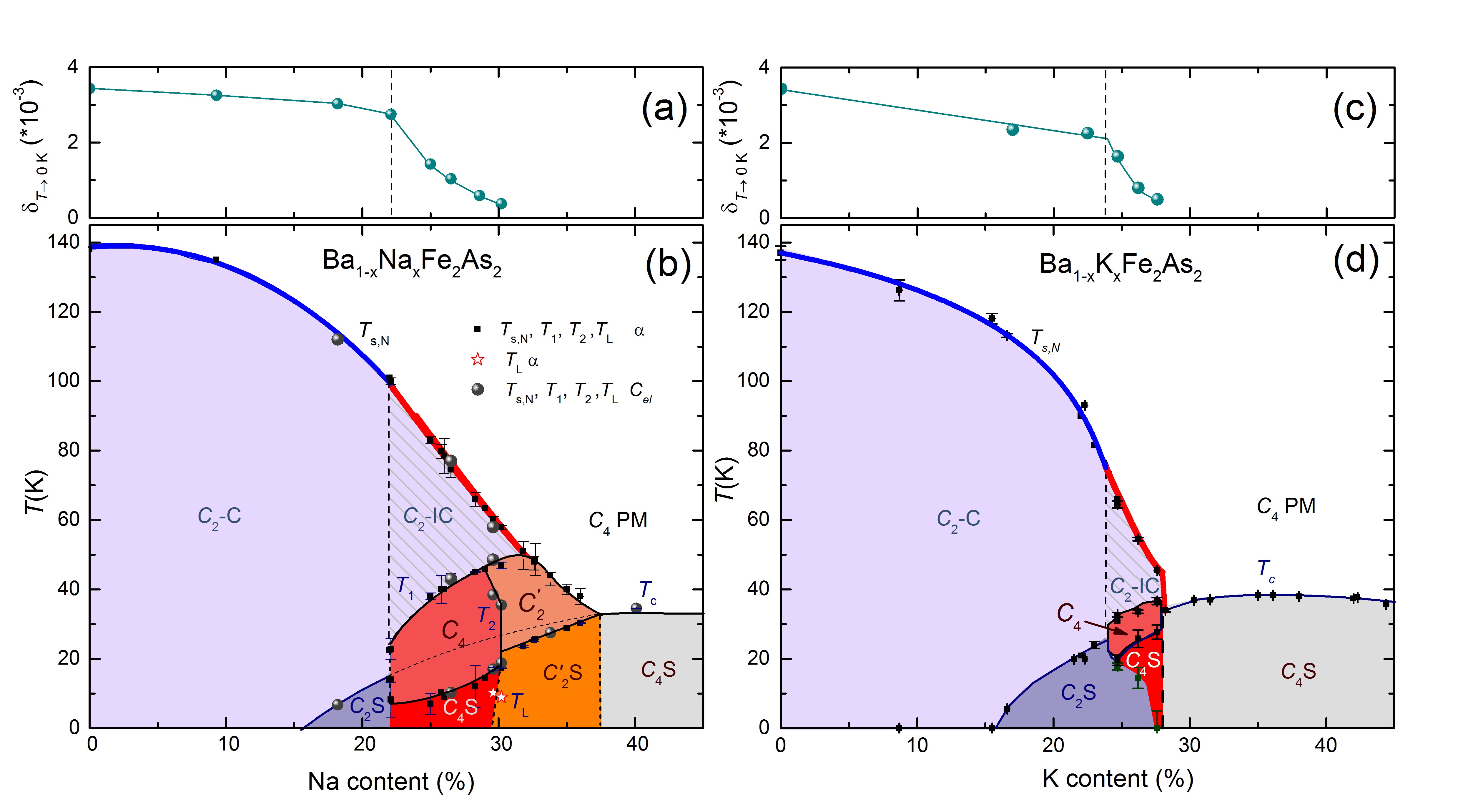}
\par\end{centering}

\protect\caption{Phase diagrams of Na- and K-doped systems. (a), (c) Extrapolated (to
T = 0) maximum orthorhombic distortion versus Na and K doping, respectively.
(b) Electronic phase diagram of Ba$_{1-x}$Na$_{x}$Fe$_{2}$As$_{2}$
obtained from thermal-expansion (squares) and specific-heat (circles)
data revealing nine different phases (see text for details). (d) Phase
diagram of Ba$_{1-x}$K$_{x}$Fe$_{2}$As$_{2}$ from Ref. \citep{Bohmer:2015ab}
for comparison. The kinks in (a) and (c), as well as the inflection
points of $T_{s,N}$ in (b) and (d), near x = 0.22 - 0.23 are interpreted
as marking a possible transition from a commensurate (C\protect\textsubscript{2}-C)
phase to an incommensurate (C\protect\textsubscript{2}-IC) phase.
This transition is indicated by the vertical dashed lines and the
color transition of the $T_{s,N}$ line from blue to red. 'S' stands
for superconductivity.}

\label{Fig3}
\end{figure*}

Single crystals of Ba$_{1-x}$Na$_{x}$Fe$_{2}$As$_{2}$ were grown
in alumina crucibles using a self-flux method with (Ba,Na): FeAs ratios
1:4 - 1:5. The crucibles were sealed in iron cylinders filled with
argon gas. After heating to 1150 - 1170 $^{0}$$C$ the furnace was
cooled down slowly at rates between 0.3 - 0.5 $^{0}$$C$ /h to minimize
the amount of flux inclusions. Near 940 - 1020 $^{0}$$C$ the furnace
was turned upside down to separate the remaining liquid flux from
the grown crystals and then cooled down to room temperature with intermediate
holds to in-situ anneal the crystals. Thermal expansion was measured
using a high-resolution home-made capacitance dilatometer \citep{Meingast1990},
which is several orders of magnitude more sensitive than traditional
diffraction techniques. Heat capacity was measured using a Physical
Property Measurement System from Quantum Design. The electronic specific
heat was obtained by subtracting an appropriate phonon background
\citep{Hardy2010,Hardy2013,Bohmer:2015ab}. Specifically, as demonstrated
for Ba$_{1-x}$K$_{x}$Fe$_{2}$As$_{2}$ \citep{Bohmer:2015ab,Hardyunpubl},
the phonon background can be approximated as the weighted sum of the
individual lattice contributions of its 'constitutents' \citep{Qiu:2001aa},
which are BaFe$_{2}$As$_{2}$ and NaFe$_{2}$As$_{2}$ for the present
case. Since there are no crystals of NaFe$_{2}$As$_{2}$, we determined
the hypothetical NaFe$_{2}$As$_{2}$ phonon background by assuming
that the electronic component at optimal doping of Na- and K-doped
\citep{Bohmer:2015ab} systems are identical. This is quite reasonable,
since both $T{}_{c}$ and the heat capacity jumps at optimal doping
are very similar in both systems. The Na content of seven single crystals
(x = 0.093(4), 0.182(2), 0.221(2), 0.283(2), 0.320(2), 0.360(3), and
0.401(4)) used for the thermal-expansion and specific-heat measurements
was accurately determined by 4-circle single crystal x-ray refinement
of a small piece of the measured crystals. The Na content of the other
crystals were interpolated between these fixed points using the SDW
transition temperature as a reference. The values of the structural
parameters from our x-ray refinement are in good agreement with previous
results \citep{Avci2013}. 

Fig. \ref{Fig1}a presents the relative thermal expansion, $\Delta L/L$,
measured along the a- and b-axes for three representative Na doping
levels. As we have demonstrated previously \citep{Boehmer2012PRB,Bohmer:2015ab},
the shorter b-axis in the low-temperature orthorhombic phase can be
obtained directly by measuring the expansion of the crystal along
the {[}110{]}\textsubscript{T} direction of the original tetragonal
cell, because in this configuration the small force from the dilatometer
detwins the crystal. The larger a-axis, on the other hand, is obtained
by combining a 'twinned' measurement (along {[}100{]}\textsubscript{T})
with the 'detwinned' data \citep{Boehmer2012PRB,Bohmer:2015ab}. The
expected orthorhombic splitting of the a- and b-lattice parameters
at the SDW transition at $T_{s,N}$ is clearly observed for all three
concentrations and reduces in magnitude with increasing Na content.
For the x = 0.265 sample, this splitting suddenly disappears, within
the accuracy of the measurements, at a first-order transition at $T{}_{1}$
= 45K, which we identify with the $C_{4}$ magnetic phase \citep{Avci2014,Bohmer:2015ab}.

In order to study the doping evolution of these transitions in greater
detail, we present in Fig. \ref{Fig1}b the orthorhombic distortion,
$\delta=(a-b)/(a+b)$, inferred from our thermal-expansion data for
a number of compositions between x = 0 and x = 0.36. We detect clear
signatures of the structural distortion associated with the SDW transition
at $T{}_{s,N}$ all the way to x = 0.36, which is considerably higher
than observed previously by neutron diffraction \citep{Avci2014,Avci2013}.
We note, however, that the orthorhombic splitting becomes extremely
small in this high-doping region (see inset of Fig. \ref{Fig1}b),
which is probably why it was missed previously. The presence of the
reentrant $C_{4}$ phase is signaled by a sudden disappearance of
$\delta$ at $T{}_{1}$, which we observe for 0.22 \ensuremath{\le}
x \ensuremath{\le} 0.29. The behavior of the lattice parameters changes
dramatically for x = 0.302, where we observe a more gradual reduction
of $\delta$ at $T{}_{1}$, indicative of a second-order transition,
followed by a previously unobserved transition at $T{}_{2}$. Upon
further doping, the transition at $T{}_{2}$ disappears and the transitions
at $T{}_{s,N}$ and $T{}_{1}$ appear to merge together. The well-known
reduction of $\delta$ at the superconducting transition in the $C_{2}$
SDW phase due to the competition between superconductivity and magnetism
\citep{Nandi2010,Boehmer2012PRB,Meingast2012} is clearly observed
for the crystal with x = 0.221, whereas the effect of superconductivity
on the in-plane lattice parameters in the $C_{4}$ phase is too small
to be seen in these curves. 

\begin{figure*}
\begin{centering}
\includegraphics[scale=0.35]{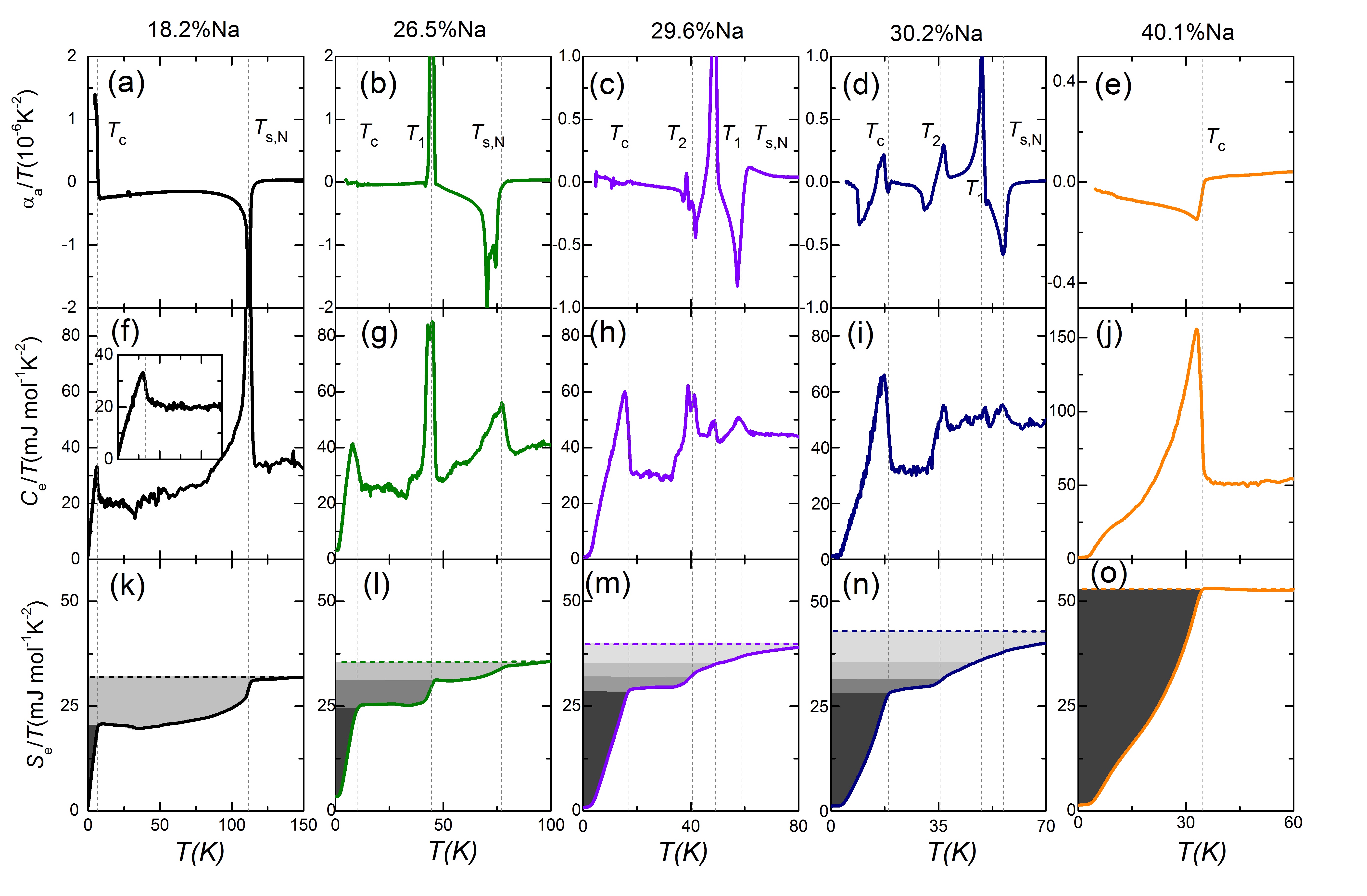}
\par\end{centering}

\protect\caption{(a)-(o) Temperature dependence of thermal expansion ($\alpha_{a}$/T)
(a-e), electronic heat capacity $C_{e}/T$ (f-j) and electronic entropy
($S_{e}/T$) (k-o) for crystals with x = 0.182, 0.265, 0.296, 0.302
and 0.401. The different shades of gray represents the step-wise reduction
of $S_{e}/T$ from the high temperature paramagnetic phase to the
low temperature superconducting state.}

\label{Fig4}
\end{figure*}

The small anomalies associated with the onset of $T{}_{c}$, as well
as the other phase transitions, are more clearly observed in the thermal-expansion
coefficients, $\alpha(T)=1/L\cdot dL(T)/dT$, for the 'twinned' and
'detwinned' directions, which are presented as $\alpha/T$ versus
T for representative Na contents in Fig.\ref{Fig2}. Fig.\ref{Fig2}a
displays data for the crystal with x = 0.182, which becomes orthorhombic
below $T{}_{s,N}$ = 112 K and superconducting below $T{}_{c}$ =
6.5 K. The clear anisotropy of the in-plane expansion below $T{}_{s,N}$,
as well as the anisotropic response at $T{}_{c}$, are indicative
of the expected orthorhombic state at this doping level. We note that
the small anisotropic tail above $T_{s,N}$ results from the small,
but finite, uniaxial pressure we apply in our dilatometer \citep{Bohmer:2015ab}.
In contrast to the behavior for x = 0.182, the anisotropy of the expansivity
vanishes nearly completely below the transition at $T{}_{1}$ for
the x = 0.265 sample (see Fig.\ref{Fig2}b), indicating the reentrant
tetragonal state below $T{}_{1}$. As expected at the onset of superconductivity,
small jump-like anomalies at $T{}_{c}$ are observed for both directions.
The behavior of the x = 0.302 crystal is more complicated (see Fig.\ref{Fig2}c).
Here the crystal clearly becomes orthorhombic at $T{}_{s,N}$, then
$\delta(T)$ decreases gradually between $T{}_{1}$ and $T{}_{2}$
(see inset of Fig.\ref{Fig1}b), but remains orthorhombic. The expansivities
for both orientations are equal below $T{}_{2}$, suggesting that
the system again enters a tetragonal state. The curves below $T{}_{c}$,
however, again exhibit an anisotropic response, suggesting that the
$C_{4}$ phase reverts back to the $C_{2}$' phase below $T{}_{c}$,
in analogy to what has been observed in K-doped BaFe$_{2}$As$_{2}$
\citep{Bohmer:2015ab}. There is an additional sharp anomaly at $T{}_{L}$=10
K for both orientations, which is however observed only upon heating,
possibly indicating another phase transition with a large thermal
hysteresis. Nearly identical behavior was observed in another crystal
with a similar composition. Our expansion data thus clearly show that
the reentrant $C_{4}$ phase exists only in a limited temperature
range between $T{}_{2}$ and $T{}_{c}$ for x = 0.302. The transitions
at $T{}_{2}$ and $T{}_{L}$ both disappear for the next higher Na
content (see Fig.\ref{Fig2}d), and this sample also clearly displays
strongly anisotropic thermal expansivities below $T{}_{s,N}$, which
is incompatible with a $C_{4}$ symmetry. The x = 0.36 crystal (Fig.
\ref{Fig2}e) exhibits only very small effects at $T{}_{s,N}$ and
$T{}_{1}$. Finally, any signature of the anomaly at $T{}_{s,N}$
has disappeared in the crystal with x = 0.401, which only has a clear
anomaly at $T{}_{c}$ = 35 K.

The transition temperatures $T{}_{s,N}$, $T{}_{c}$, $T{}_{1}$,
$T{}_{2}$ and $T{}_{L}$ obtained by the thermal -expansion data
shown in Figs. \ref{Fig1} and \ref{Fig2} allow us to construct a
detailed phase diagram (see Fig.\ref{Fig3}b). Here, we have also
included the transition temperatures extracted from the heat-capacity
data (see Fig.\ref{Fig4}). The phase diagram exhibits a remarkable
degree of complexity, with a surprising number of additional (other
than the usual $C_{2}$-magnetic and superconductivity) phases emerging
as magnetism is suppressed by Na doping. We note that these phases
appear to emerge at the point where $T{}_{s,N}$ changes curvature
from concave to convex near x = 0.22. This change is indicated by
the changing color of the line from blue to red. The doping dependence
of the extrapolated zero-temperature orthorhombic distortion of the
$C_{2}$ phase (see Fig. \ref{Fig3}a) illustrates this change even
more clearly, with a very distinct kink near x = 0.22. We interpret
the inflection point of $T{}_{s,N}$(x) as a sign for a commensurate-to-incommensurate
transition as expected in a simple mean-field SDW picture \citep{Vorontsov2010,RicePRB1970,KulikovSPU1984}.
Previously, clear evidence for incommensurability has only been reported
in electron doped BaFe$_{2}$As$_{2}$ \citep{Pratt2011,Bonvilleepl,Luo:2012aa}.
Since we do not observe a splitting of the $T{}_{s,N}$ line into
two transitions above x = 0.22, we postulate the vertical dashed line
to indicate the proposed commensurate-to-incommensurate transition.
Such a vertical line implies a first-order transition, evidence of
which is provided by the jumps of both $T{}_{1}$ and $T{}_{c}$ at
x = 0.22. In Fig. \ref{Fig3}c and d we compare the present results
to those of K-doped BaFe$_{2}$As$_{2}$ \citep{Bohmer:2015ab}. Similar
to the Na-doped system, we also find an inflection point in its phase
diagram (see Fig. \ref{Fig3} d), as well as a kink in the T = 0 orthorhombic
distortion (see Fig. \ref{Fig3}c), at a K content at which the $C_{4}$
phase emerges (see Fig. \ref{Fig3}c and d). This strongly suggests
that these are both common features in hole doped BaFe$_{2}$As$_{2}$.
In Ba$_{1-x}$Na$_{x}$Fe$_{2}$As$_{2}$we observe, in addition to
the magnetic $C_{4}$ phase, a previously unobserved phase (labeled
$C_{2}$' in Fig. \ref{Fig3} b), which has a reduced, but finite,
orthorhombic distortion. A similar phase in not observed in the K-doped
system. Although we can not examine the microscopic order in this
phase using our macroscopic probes, the smooth doping variations of
both $T{}_{s,N}$ and $\delta$, suggest that this phase is probably
also of magnetic origin, although some kind of charge \citep{Gastiasoro:2015}
order cannot be excluded. Preliminary \textgreek{m}SR measurements
on a crystal with x=0.33 provide evidence for a magnetic $C_{2}$'
phase \citep{Malletunpubl}. Detailed investigations of the magnetic
structure using e.g. neutron scattering are highly desirable once
large enough crystals become available.

In order to gain more insight into the different phases, we present
the electronic heat capacity for several Na concentrations in Fig.\ref{Fig4}
together with thermal expansion of the a-axis for comparison. As demonstrated
in Fig. \ref{Fig3} and \ref{Fig4}, the transition temperatures from
the heat capacity (solid gray circles in Fig.\ref{Fig3}) closely
match those from the thermal expansion. With increasing Na-doping
the step-like anomalies in $C_{e}/T$ associated with superconductivity
generally increase in size, whereas the anomalies associated with
the magnetic transitions weaken, indicating the well-known competition
between magnetism and superconductivity in the Fe-based systems \citep{Nandi2010,Meingast2012,Boehmer2012PRB,Bohmer:2015ab}.
This trend is made even more transparent in Fig.\ref{Fig4} k-o, where
we plot $S_{e}/T=(\intop C_{e}/TdT)/T$, i.e. the electronic entropy
divided by T, which for a Fermi liquid is expected to be constant.
Upon entering the $C_{4}$ phase at 45 K for the x = 0.265 Na sample
we observe a particularly large additional reduction of $S_{e}/T$
at $T{}_{1}$, which is more prominent than the anomaly at $T{}_{s,N}$,
and apparently results in a large suppression of $T{}_{c}$ and the
condensation energy (equal to the black shaded area in Fig. \ref{Fig4}
k)-o)) in the $C_{4}$ phase. This highlights the much stronger competition
of superconductivity with the double-Q $C_{4}$ magnetic phase than
with the usual magnetic $C_{2}$ phase, which was also observed in
Ba$_{1-x}$K$_{x}$Fe$_{2}$As$_{2}$ \citep{Bohmer:2015ab}. However,
in contrast to Ba$_{1-x}$K$_{x}$Fe$_{2}$As$_{2}$ \citep{Bohmer:2015ab},
we find no evidence for a reemergence of the usual stripe-type $C_{2}$
phase below $T{}_{c}$. For the crystals with x = 0.296 and 0.302,
the largest (non superconducting) anomalies in $C_{e}/T$ and $S_{e}/T$
occur not at $T{}_{1}$, but rather upon entering the $C_{4}$ phase
at $T{}_{2}$. Interestingly, the $S_{e}/T$ plot for both these samples
(Fig. \ref{Fig4} m and n) provide evidence for a pseudogap-like behavior
above $T{}_{2}$ - i.e. a gradual loss of density-of-states as the
temperature is lowered. The competition of superconductivity with
the $C_{2}$' phase appears to be much weaker than with the $C_{4}$-magnetic
phase, as evidenced by the increase of the superconducting condensation
energy , as well as the rise of $T{}_{c}$ seen in Fig. \ref{Fig3}
within the $C_{2}$' phase. Finally, we note that the negligible residual
$C_{e}/T$ values of all of our samples (except for x = 0.265) demonstrate
that our samples are of high quality and that doping away from the
FeAs layer does not introduce pair breaking, as it does in Co-doped
BaFe$_{2}$As$_{2}$ \citep{Hardy2010a}.

In summary, our detailed thermodynamic studies of Ba$_{1-x}$Na$_{x}$Fe$_{2}$As$_{2}$
show that the phase diagram of this system exhibits a surprising degree
of complexity. As stripe-type magnetism is suppressed by Na-doping,
two additional magnetic phases emerge, which coexist and compete with
superconductivity. The emergence of these additional phases is shown
to be possibly triggered by a doping-induced commensurate-incommensurate
transition near x = 0.22, which would provide further evidence for
electronic itinerancy in these systems. There are many similarities
between the phase diagrams of K- and Na-doped BaFe$_{2}$As$_{2}$,
and the differences are likely related to chemical pressure, since
our previous studies on the K-doped system have shown that the phase
boundaries are extremely pressure dependent \citep{Bohmer:2015ab}.
Importantly, the presently observed complexity of these phase diagram
suggests a high degree of degeneracy of several energy scales as the
optimally-doped state is approached, which may also be related to
the superconducting pairing mechanism. 

We acknowledge fruitful discussions with Christian Bernhard, Markus
Braden, Rafael Fernandes, Maria Gastiasoro, Benjamin Mallett, Jörg
Schmalian, and Florian Waßer.

\bibliographystyle{apsrev4-1}
\addcontentsline{toc}{section}{\refname}\bibliography{NaBa122}

\end{document}